\documentclass{aa}
\usepackage{graphicx}
\begin{document}

\title{The DIRECT project: Catalogs of stellar objects in nearby
galaxies. II. Eastern arm and NGC 206 in M31.\thanks{Tables
1$-$5 are only available in electronic form at the CDS via anonymous
ftp to {\tt cdsarc.u-strsbg.fr} (130.79.125.5) or via {\tt
http://cdsweb.u-strsbg.fr/Abstract.html} They are also available
through {\tt anonymous ftp} on {\tt cfa-ftp.harvard.edu}, in {\tt
pub/kstanek/DIRECT} directory. 
}}
\author{B.~J. Mochejska\inst{1}
\and J. Kaluzny\inst{1}
\and K.~Z. Stanek\inst{2}
\and D.~D. Sasselov\inst{2}}

\institute{
Copernicus Astronomical Center, Bartycka 18, 00-716 Warszawa, Poland
\email{mochejsk, jka@camk.edu.pl}
\and
Harvard-Smithsonian Center for Astrophysics, 60 Garden St.,
Cambridge, MA~02138 
\email{kstanek, dsasselov@cfa.harvard.edu}
}

\offprints{B.~J. Mochejska, \email{mochejsk@camk.edu.pl}}

\date{Received ...., 2001; accepted ...., 2001}

\abstract{ DIRECT is a project to directly obtain the distances to two
important galaxies in the cosmological distance ladder, M31 and M33,
using detached eclipsing binaries and Cepheids. As part of our search
for these variables, we have obtained photometry and positions for
thousands of stellar objects within the monitored fields, covering an
area of 557.8 arcmin$^2$. In this research note we present the
equatorial coordinates and $BVI$ photometry for 26\,712 stars in the
M31 galaxy, along the eastern arm and in the vicinity of
the star forming region NGC206. 
 }

\titlerunning{Catalog of stellar objects in M31}

\maketitle

\section{Introduction}

Starting in 1996 we undertook a long term project, DIRECT, to obtain
the distances to two important galaxies in the cosmological distance
ladder, M31 and M33. These ``direct'' distances will be obtained by
determining the distance of Cepheids using the Baade-Wesselink method
and by measuring the distance to detached eclipsing binaries
(DEBs).

As the first step of the DIRECT project we have searched for DEBs and
new Cepheids in M31 and M33. In the M31 galaxy we have analyzed five
$11\arcmin\times11\arcmin$ fields, A$-$D and F (Kaluzny et
al. \cite{pap1}, \cite{pap4}; Mochejska et al. \cite{pap5}; Stanek et
al. \cite{pap2}, \cite{pap3}; hereafter Papers I, IV, V, II, III). A
total of 410 variables, mostly new, were found: 48 eclipsing binaries,
206 Cepheids and 156 other periodic, possible long-period or
non-periodic variables. In the first paper of the series of stellar
catalogs we have presented the catalog of stars in the central part of
M33 (Macri et al. \cite{lmm}). In this research note we present the
equatorial coordinates and $BVI$ photometry for 26\,712 stars detected
in the M31 galaxy within the monitored area of 557.8 arcmin$^2$.

\begin{figure}
\resizebox{\hsize}{!}{\includegraphics{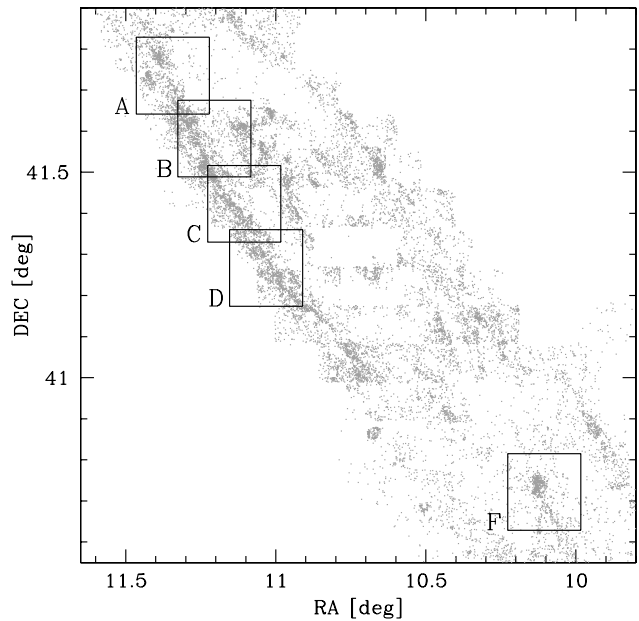}}
\caption{Location of the fields A$-$D and F observed in M31.}
\label{fig:xy}
\end{figure}

\begin{figure}
\resizebox{\hsize}{!}{\includegraphics{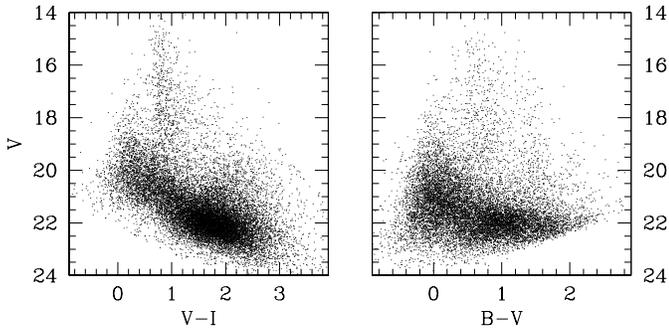}}
\caption{V/V-I and V/B-V color magnitude diagrams.}
\label{fig:cmd}
\end{figure}

\section{Observations and data reduction}
We have observed five fields, A$-$D, concentrated on the rich spiral
arm, and F, containing the giant star formation region known as NGC
206 (Fig. \ref{fig:xy}). The center ($\alpha_{2000}$, $\delta_{2000}$)
coordinates of the fields were: A (11\fdg34, 41\fdg73); B (11\fdg20,
41\fdg59); C (11\fdg10, 41\fdg42); D (11\fdg03, 41\fdg27); F
(10\fdg10, 40\fdg72);

M31 was primarily observed in 1996 with the 1.3~m McGraw-Hill
Telescope at the Michigan-Dartmouth-MIT (MDM) Observatory. We used the
front-illuminated, Loral $2048^2$ CCD ``Wilbur'' (Metzger, Tonry \&
Luppino \cite{metz}). Data for M31 were also obtained, mostly in 1997,
with the 1.2~m telescope at the F. L. Whipple Observatory (FLWO),
where we used ``AndyCam'' (Szentgyorgyi et al. \cite{sze}), with a
thinned, back-illuminated, AR coated Loral $2048^2$ pixel CCD. The
pixel scale was essentially the same at both telescopes, $0.32$ arcsec
pixel$^{-1}$, giving a field of view of roughly $11\times11$
arcmin$^2$.

For the full description of the applied data reduction, calibration
and astrometry procedures the reader is referred to Paper I. Here we
present only a very brief summary. Stellar profile photometry was
extracted using the Daophot/Allstar package (Stetson \cite{ste1}). The
transformation of instrumental magnitudes to the standard system was
based on 18 standard stars (Landolt \cite{lan}) observed on the night
of 1996 September 14/15 at MDM. The residuals in $V$, $V-I$ and $B-V$
showed no overall offsets and no dependence on color (Fig. 2 in Paper
I). A comparison with the Magnier et al. (\cite{mag}) photometry
showed very good agreement in $V$ (average $V-V_{Ma92}$=0.013 for
stars with $V<20$) and a strong trend in $V-I$ residuals with the
$V-I$ color (Fig. 4 in Paper I). We have obtained independent
calibrations at MDM on the night of 1996 October 2/3 with the
Charlotte 1024$^2$ CCD (field B) and at FLWO on the night of 1997
October 9/10 (fields C, D, F). The offsets in $V$ and $V-I$,
respectively, were 0.040 and 0.016 in field B, 0.012 and 0.024 in
field C, -0.014 and 0.047 in field D, -0.020 and 0.057 in field
F. Apart from the offsets, we did not see anything resembling the
strong trend in the $V-I$ residuals, seen in the comparison with the
photometry of Magnier et al. (\cite{mag}).  This discrepancy certainly
deserves further attention. To check the internal consistency of our
calibration, we have compared the photometry in the overlapping
regions between the fields. The offsets in $V$ and $I$, respectively,
were 0.022 and 0.018 between fields A and B, 0.034 and 0.024 between B
and C, and -0.063 and -0.040 between C and D. The offset in $B$
between fields C and D was 0.007.

The transformation from rectangular to equatorial coordinates was
derived using stars from the list published by Magnier et al. (\cite{mag})
for fields A$-$D, and the USNO-A2 catalog (Monet et al. \cite{mon})
for field F.

\section{The catalog}
In Figure \ref{fig:cmd} we plot the $V/V-I$ and $V/B-V$
color-magnitude diagrams (CMDs) from the combined A$-$D and F field
catalogs. In the $V/V-I$ CMD (left panel), stars near $V\sim22$ mag
and $V-I\sim1.8$ represent the top of the evolved red giant
population. The vertical strip of stars with $0.6<V-I<1.2$ and $V<20$
are Galactic foreground stars. Stars bluer than $V-I<0.6$ are the
upper main sequence stars in M31. In the $V/B-V CMD$ (right panel),
the most prominent feature is the upper main sequence at $B-V\sim0$.
The Galactic foreground stars are also present, between 0.4 and 1.0 in
$B-V$.

In Tables 1$-$5, we present the catalogs for fields A$-$D and F, with
the equatorial coordinates and photometry in $VI$ (fields A and B) and
$BVI$ (C, D, F). For each star we list its ID, $\alpha_{2000}$,
$\delta_{2000}$ equatorial coordinates, standard V, I and B magnitudes
with their respective errors, and the Stetson variability index J$_S$
(Stetson \cite{ste2}).  The IDs, based on the equatorial coordinates,
are in the format D31Jhhmmss.s+ddmmss.s. The first three correspond to
$\alpha$, expressed in hours (hhmmss.s), the last three to $\delta$ in
degrees (ddmmss.s).

\begin{acknowledgements}
BJM and JK were supported by the Polish KBN grant 2P03D003.17. BJM was
also supported by the Foundation for Polish Science stipend for young
scientists and the Polish KBN grant 2P03D025.19 and JK by the NSF
grant AST-9819787. DDS acknowledges support from the Alfred P. Sloan
Foundation and from NSF grant No. AST-9970812.
\end{acknowledgements}

\end{document}